\documentclass[showpacs,twocolumn,aps,amssymb,preprintnumbers,
superscriptaddress,prc,floatfix]{revtex4-1}
\usepackage{epsfig,dcolumn}
\usepackage{graphicx,bm}
\usepackage{bm,color}
\newcommand{\be}{\begin{equation}}
\newcommand{\ee}{\end{equation}}
\newcommand{\ba}{\begin{eqnarray*}}
\newcommand{\ea}{\end{eqnarray*}}

\begin{document}

\title{Collectivity in the light Xenon isotopes: A shell model study}

\author{E. Caurier}
\affiliation{IPHC, IN2P3-CNRS et Universit\'e Louis Pasteur, 
F-67037 Strasbourg, France}

\author{F. Nowacki}
\affiliation{IPHC, IN2P3-CNRS et Universit\'e Louis Pasteur, 
F-67037 Strasbourg, France}

\author{A. Poves}
\affiliation{Departamento de F\'isica Te\'orica e IFT-UAM/CSIC, 
Universidad Aut\'onoma de Madrid,  E-28049 Madrid, Spain}

\author{K. Sieja}
\affiliation{IPHC, IN2P3-CNRS et Universit\'e Louis Pasteur, 
F-67037 Strasbourg, France}

\begin{abstract}

The lightest Xenon isotopes are studied in the framework of the
Interacting Shell Model (ISM). The valence space comprises all the orbits lying
between the magic closures N=Z=50 and N=Z=82.  
The calculations produce collective deformed structures
of triaxial nature that encompass nicely the known experimental data. Predictions are
made for the (still unknown) N=Z nucleus  $^{108}$Xe. The results are 
interpreted in terms of the competition between the quadrupole correlations 
enhanced by the pseudo-SU(3) structure of the positive parity orbits and the
pairing correlations brought in by the 0h$_{11/2}$ orbit. We have studied as well  the 
effect of the excitations from the $^{100}$Sn core on our predictions. 
 We show that the backbending in this region is due
to the alignment of two particles in the 0h$_{11/2}$ orbit. In the N=Z case, one neutron and
one proton align to J=11 and T=0. In $^{110,112}$Xe the alignment  begins in the J=10 T=1
channel and it is dominantly of neutron neutron type. Approaching the band termination the  
alignment of a neutron and a proton to J=11 and T=0 takes over.
In a more academic mood, we have explored  
the role of the isovector and  isoscalar pairing correlations on the structure on the
yrast bands of   $^{108,110}$Xe and examined the role of the isovector and isoscalar
pairing condensates in these N$\sim$Z nuclei.

\end{abstract}


\date{\today}
\maketitle

\section{Introduction}

The lightest Xenon isotopes have been recently explored down to $^{110}$Xe, only two neutrons
away of the would-be N=Z nucleus  $^{108}$Xe. In a very demanding experiment carried out at 
the University of Jyv\"askyl\"a, using the recoil-decay tagging  technique  \cite{sand07}, the excitation 
energies of the lowest members of the yrast band were measured. The experimental spectrum shows 
remarkable rotational-like features. The authors of this work argue that such a behavior is unexpected
and invoke a large depletion of the doubly magic closure in  $^{100}$Sn to explain it.
However, it is well known that nuclei with six neutrons and four  protons on top of a N=Z doubly 
magic core develop collective features that can be explained microscopically without resorting to
extensive core excitations. The same is true, and even more accentuated, for the nuclei with four
neutrons and four protons outside the corresponding N=Z cores. Well documented cases are
$^{26}$Mg and  $^{24}$Mg in the $sd$-shell (core of $^{16}$O) and    $^{50}$Cr and  $^{48}$Cr
in the $pf$-shell (core of $^{40}$Ca). In both cases the closures correspond to major shells of
the harmonic oscillator.  The next magic numbers are originated by the spin-orbit interaction,
giving rise first to the $^{56}$Ni core and to the nuclei  $^{66}$Ge  and  $^{64}$Ge, that show also collective
features, even if less prominent than in the other cases. Indeed, the next N=Z  spin-orbit closure
is  $^{100}$Sn,  and the replicas of the lower mass isotopes are  $^{110}$Xe  and $^{108}$Xe.

 For  $^{26}$Mg and  $^{24}$Mg, Elliott's SU(3) \cite{elliott56}  gives the explanation of
 the onset of collectivity. Both nuclei are well deformed and triaxial, due to the predominance of the
 quadrupole-quadrupole part of the effective nuclear interaction, either in the limit of degenerate
 $sd$-shell orbits (SU(3) proper) or in the limit of very large spin-orbit splitting (quasi-SU3) \cite{zrpc95}).
 $^{48}$Cr and  $^{50}$Cr have became the paradigm of deformed nuclei amenable to a fully microscopic
 description in the laboratory frame via the interacting shell model (ISM) \cite{cau95,cau96,lenzi96}. The
 underlying coupling scheme in this region is quasi-SU3. 
 
 As we shift to the spin-orbit driven shell closures,
 the situation varies again. For instance, the quasi-degenerate orbits above  $^{56}$Ni, 1p$_{3/2}$,
  0f$_{5/2}$, and 1p$_{1/2}$, form a pseudo-SU3 sequence \cite{psu3}, with principal quantum number
  p=2. The physical valence space in this region includes also the   0g$_{9/2}$ orbit, the down-going
  intruder of opposite  parity, which couples to the pseudo-SU(3) block through the off-diagonal pairing
  matrix elements. {\it Mutatis mutandis} the same physics occurs in the region of very proton rich nuclei
  beyond  $^{100}$Sn.
  
  The plan of the paper is as follows: In section II we describe the framework of the calculations; valence 
  space, effective interaction, etc. We also deal with the group theoretical issues of the pseudo-SU(3) scheme,
  and advance some of its predictions. In section III we discuss the results for   $^{110,112}$Xe, comparing
  with the experimental results, and we make predictions for  $^{108}$Xe. Then we study the influence of the core
  excitations on the different observables and  finally we analyze the mechanisms that produce the backbending 
  of  the yrast bands of both  $^{108}$Xe  and $^{110}$Xe. In section IV  we examine in detail
  some, much debated, aspects of the physics of the N=Z nuclei. Among others, the role of the isovector and 
   isoscalar L=0 pairing components of the interaction. In section V we gather our conclusions.

\section{Valence space. Effective interaction}

The valence space is the one comprised between the magic closures N=Z=50 and N=Z=82. This
means that we have an inert core of $^{100}$Sn. The space contains two doubly magic nuclei 
$^{132}$Sn and $^{164}$Pb.

We use the effective interaction GCN50:82 which is obtained from a realistic G-matrix \cite{morten95}
based upon the Bonn-C potential \cite{bonnc}.  Taking the G-matrix as the starting point, different combinations
of two body matrix elements are fitted  to a large set of experimental
excitation energies in the region (about 400 data points from 80 nuclei).
 As can be seen in Table \ref{tab:rms} the modifications of the
 monopole hamiltonian are, as usual, the ones that improve most the agreement with the data.
 Subsequent modifications of the pairing and multipole hamiltonians bring the root mean square deviation
 to a very satisfactory value of 110~keV \cite{gcn5082}.

\begin{table}[h]
\caption{\label{tab:rms}Evolution of the rms deviation with the corrections brought to the realistic interaction}
\begin{tabular*}{\linewidth}{@{\extracolsep{\fill}}lc}
\hline\hline
Corrections  &rms(MeV)\\ \hline
G (none)                   &1.350\\
+monopole             &0.250\\
+pairing                  &0.180\\
+multipole                    &0.110\\
\hline\hline
\end{tabular*}
\end{table}

In this valence space the dimensions grow rapidly reaching O(10$^{10}$) in some of the calculations presented here.
In spite of this, one can have an idea of the kind of results that can be achieved in this model space in the limit
of pure pseudo-SU3 symmetry.  The Nilsson-like orbits of pseudo-SU3 corresponding to the principal quantum
number $p$ have intrinsic quadrupole moments given by  \cite{rmp},
$q(p,\chi,k) =(-2p -3\chi) b^2$, where $\chi$ can take integer  values between 0 and {\it p},
$k=\pm(\frac{1}{2}  \ldots  \frac{1}{2}+\chi)$ and b is the harmonic oscillator length parameter. The total quadrupole moment is the sum of the contributions of
all the valence protons and neutrons with the corresponding effective charges. The orbits are filled orderly,
starting from $\chi$=0 or  $\chi$={\it p} depending on which choice gives the largest total intrinsic quadrupole
moment in absolute value. This is so because in this scheme the correlation energy is proportional to Q$^2_0$.

 The four valence protons of the Xenons can adopt several degenerate configurations; 

\begin{quote}
 $(\chi=0,k=\pm\frac{1}{2})^2$$(\chi=1,k=\pm\frac{1}{2})^2$;
 
$(\chi=0,k=\pm\frac{1}{2})^2$$(\chi=1,k=\pm\frac{3}{2})^2$ and

 $(\chi=0,k=\pm\frac{1}{2})^2$$(\chi=1,k=\pm\frac{1}{2})$\\$(\chi=1,k=\pm\frac{3}{2})$,
\end{quote}

\noindent
 leading to 
K=0 and K=2. Even a small K-mixing produces triaxiality and we should expect it to show up in the
experiments and in the realistic calculations.

 Moving to  the neutron side, the model predicts that  triaxiality should be larger in $^{108}$Xe
 than in $^{110}$Xe because of the neutron contribution. In addition,    
   it turns out that for {\bf p}=3 the contribution to the intrinsic quadrupole
 moments of the valence neutrons in excess of six is zero. The predicted values for the light Xenon
  isotopes are: Q$_0$($^{108}$Xe)=~210~e~fm$^2$ and Q$_0$($^{110-116}$Xe)=~225~e~fm$^2$.
  This corresponds to B(E2)(2$^+$$\rightarrow$0$^+$) of 870~e$^2$~fm$^4$ and
  1000~e$^2$~fm$^4$ respectively. Therefore, the increase of collectivity that is seen experimentally 
  toward mid-shell, reaching a maximum in $^{120}$Xe with 16 valence neutrons cannot be explained 
  in this scheme.  To get more quadrupole collectivity we need to enlarge the valence space, and the
  more conservative choice is to include the 1f$_{7/2}$ neutron orbit, which is the  quasi-SU3 partner of the 
  0h$_{11/2}$. Assuming that the valence neutrons in excess of six occupy the lowest quasi-SU3
  intrinsic orbits corresponding to  {\bf p}=5, the quadrupole moments will keep increasing up to 
  $^{120}$Xe as demanded by the experimental data.

  \begin{table}
\caption{B(E2)(0$^+$$\rightarrow$ 2$^+$) for the Xenon isotopes (in e$^2$b$^2$)}
\label{tab:xe_quasi}
\begin{tabular*}{\linewidth}{@{\extracolsep{\fill}}cccccc}
    \hline \hline 
A= & 114 & 116 & 118 & 120      & 122 \\ [1pt]
\hline 
SU3 & 1.05&  1.27    &   1.52   & 1.60  &  1.62     \\ [2pt]
EXP   & 0.93(6) & 1.21(6) & 1.40(7) & 1.73(11) &  1.40(6)  \\ [2pt]
       \hline \hline
    \end{tabular*}
\end{table}
  
  We compare in Table \ref{tab:xe_quasi} the predictions of this simple model with the experimental
  results. The agreement is astonishingly good. We have made exploratory calculations for $^{112}$Xe 
  finding an increase of the B(E2) from 1130~e$^2$~fm$^4$      in the pseudo-SU3 valence
  space to 1460~e$^2$~fm$^4$ in the space corresponding to the proposed  pseudo+quasi-SU3 scenario.

\section{Spectroscopy of the light Xenon isotopes}

\subsection{$^{110}$Xe}

\begin{figure}[h]
 \begin{center}
\includegraphics[width=0.4\textwidth]{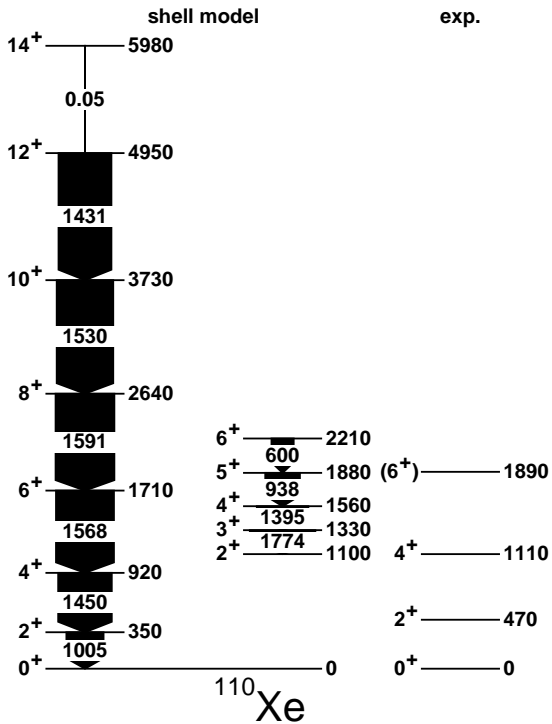}
\caption{\label{fig:xe110}(color online) The energy levels of  $^{110}$Xe.}
  \end{center}
\end{figure}

 Let's start with the results for $^{110}$Xe. The calculated energy levels are shown in Fig. \ref{fig:xe110} in
 comparison with the experimental results from \cite{sand07}. The agreement is quite satisfactory even if the
 theoretical calculation predict a moment of inertia slightly larger than the experimental one. We have checked that it is possible to obtain perfect agreement for the excitation energies, without changing the quadrupole properties
 of the band, just increasing  $\sim$~20\% the isovector pairing channel of the effective interaction. Nevertheless this  
 residual discrepancy does not justify to make modifications in the effective interaction GCN50:82 which was
 designed to give good spectroscopy in the full r4h valence space.
 Notice that the rotational   features of the shell model
 calculation persist till the backbending that occurs at J=14. In addition, the calculation  produces a $\gamma$
 band which will be analyzed  below.

   In Table \ref{tab:xe110y} we collect the  quadrupole properties of the yrast band. They are compatible
   with a deformed intrinsic state with constant intrinsic quadrupole moment Q$_0$=200~efm$^2$, corresponding
   to $\beta$=0.16, not far from the expectations of the pseudo-SU3 model. Beyond J=12 the alignment
    regime takes over, as we shall analyze  in detail in the next sections, 
   and the   rotation is no longer collective.

\begin{table}
\caption{\label{tab:xe110y} Properties of the yrast band of $^{110}$Xe 
 (energies in MeV, Q's in efm$^2$ and BE2's in e$^2$fm$^4$)}
\begin{tabular*}{\linewidth}{@{\extracolsep{\fill}}cccrrrrr}

\hline \hline \\ [2pt]
J & E* & E$_{\gamma}$ & BE2 & Q$_{sp}$ & Q$_0$      & Q$_0$  & $\beta$\\
  &    &              &       &       & (BE2)    & (Q$_{sp}$) \\ [2pt]  \hline \\ [2pt]
 2$^+$  & 0.35 & 0.35 & 1005 & -62  & 225 & 217 &  0.17 \\ [2pt]
 4$^+$  & 0.92 & 0.57 & 1450 & -78  & 226 & 215 &  0.17 \\ [2pt]
 6$^+$  & 1.71 & 0.79 & 1568 & -83  & 224 & 208 &  0.17 \\ [2pt]
 8$^+$  & 2.64 & 0.94 & 1591 & -87  & 220 & 207 &  0.17 \\ [2pt]
10$^+$  & 3.73 & 1.09 & 1530 & -86  & 213 & 198 &  0.17 \\ [2pt]
12$^+$  & 4.95 & 1.22 & 1431 & -85  & 204 & 191 &  0.16 \\ [2pt]
14$^+$  & 5.98 & 0.99 & 0.05 & -126 &    &  &       \\ [2pt]
16$^+$  & 6.63 & 0.69 & 111  & -125 &    &  &       \\ [2pt]
18$^+$  & 7.51 & 0.88 & 1184 & -130 &    &  &       \\ [2pt]
20$^+$  & 8.51 & 1.00 & 1043 & -134 &    &  &       \\ [5pt]
       \hline \hline
    \end{tabular*}
\end{table}

 We can now turn back to the $\gamma$ band. Besides the characteristic sequence of energy
 levels shown in Fig. \ref{fig:xe110}  the quadrupole properties of the band, gathered in Table \ref{tab:xe110g},
 strongly suggest that  the  $\gamma$ band of $^{110}$Xe has K=2.
 In the first place, the intrinsic quadrupole moment, extracted from
 the spectroscopic quadrupole moments or from the B(E2)'s, assuming K=2, is fairly constant and very
 close to that of the yrast band.  The more so if we realize that 
Q($2^+_{\gamma}$)=-Q($2^+_{y}$) and Q(3$^+$)$\approx$0  as it should be the case
if the $\gamma$ band could be labeled by K=2.

  The question now is whether $^{110}$Xe it is a triaxial nucleus or not.
 For that, a certain amount of K-mixing is required. In fact,  in the limit of pure pseudo-SU3 symmetry 
 the mixing is zero and this  would not be the case.  In some models \cite{davi}, the amount of triaxiality, $\gamma$
is derived from the ratio: 

\begin{equation}
\frac{BE2(2^+_{\gamma} \rightarrow 2^+_{y})}{BE2(2^+_{\gamma} \rightarrow 0^+_{y})}
\end{equation}

In our case this corresponds to a value $\gamma$=20$^o$. 
\begin{table}
\caption{Properties of the $\gamma$ band of $^{110}$Xe
 (energies in MeV, Q's in efm$^2$ and BE2's in e$^2$fm$^2$)}
\label{tab:xe110g}
\begin{tabular*}{\linewidth}{@{\extracolsep{\fill}}cccrrrrr}
    \hline \hline \\ [2pt]
J & E* & E$_{\gamma}$ & BE2 & Q$_{sp}$ & Q$_0$(BE2)      & Q$_0$(Q$_{sp}$)  & $\beta$\\ [2pt]
\hline \\ [2pt]
2$^+_2$ & 1.10 &      &      & +61  &     &  214   &   0.17    \\ [2pt]
3$^+$   & 1.33 & 0.23 & 1774 & -1.3 &     &     &       \\ [2pt]
4$^+_2$ & 1.56 & 0.23 & 1395 & -38  & 219 & 261 &  0.18 \\ [2pt]
5$^+$   & 1.88 & 0.32 & 938  & -54  & 217 & 234 &  0.17 \\ [2pt]
6$^+_2$ & 2.21 & 0.33 & 600  & -74  & 209 & 259 &  0.18 \\ [5pt]
       \hline \hline
    \end{tabular*}
\end{table}
As we have already mentioned, the valence space r4h contains a pseudo-SU3 triplet
plus the intruder orbit 0h$_{11/2}$. When we remove 
the intruder orbit  from the space, 
the moment of inertia of the nucleus gets reduced by 30\%, the backbending is suppressed,
the triaxiality is reduced to $\gamma=12^o$; and the magnetic moments become fully 
consistent with the rotational model up to J=20. The changes in the E2 transition probabilities
and quadrupole moments below the backbending region are negligible.

\subsection{$^{112}$Xe}

    The shell model description of this isotope is a real challenge because the dimension of the basis in the full
    space calculation (number of M=0 Slater determinants) is $\sim$~10$^{10}$ (exactly 9324751339). The results
     are gathered in Table~\ref{tab:xe112} and compared with the available experimental data \cite{smith01}. 
     For the lowest part of the yrast band they resemble very much
     to  those of  and the agreement with the experimental excitation energies is even better.
       Nevertheless,
      when entering in the backbending region, which is predicted by the calculation at the right spin, J=10,
      the accord deteriorates somehow. As we have discussed in the previous 
      sections this can be a  manifestation of  the limitations of our valence for the description of
       the heavier Xenon isotopes.  The backbending  
       corresponds to the alignment of two neutrons in the   0h$_{11/2}$ orbit
        as in the lighter isotope $^{110}$Xe.
       This change of structure 
      is clearly seen in the drastic reduction of the B(E2) of the 10$^+$$\rightarrow$8$^+$ transition,
       which is simultaneous
      with an increase of the spectroscopic quadrupole moment of the 10$^+$  state.
            
\begin{table}[h]

\caption{\label{tab:xe112} Properties of the yrast band of $^{112}$Xe 
 (energies in MeV, Q's in efm$^2$ and BE2's in e$^2$fm$^4$)}
\begin{tabular*}{\linewidth}{@{\extracolsep{\fill}}cccrrrrr}

\hline \hline \\ [2pt]
J & E* & E$_{exp}$ & BE2 & Q$_{sp}$ & Q$_0$      & Q$_0$  & $\beta$\\
  &    &              &       &       & (BE2)    & (Q$_{sp}$) \\ [2pt]  \hline \\ [2pt]
 2$^+$  & 0.38 & 0.46 & 1063 & -62  & 217 & 231 &  0.17 \\ [2pt]
 4$^+$  & 1.00 & 1.12 & 1560 & -75  & 206 & 236 &  0.17 \\ [2pt]
 6$^+$  & 1.82 & 1.91 & 1727 & -76  & 190 & 236 &  0.17 \\ [2pt]
 8$^+$  & 2.79 & 2.78 & 1783 & -74  & 176 & 232 &  0.17 \\ [2pt]
 10$^+$  &  3.72 & 3.55 & 600 & -97 &  &  &   \\ [2pt]
 12$^+$  & 4.20 & 4.47 & 1471 & -118  &  &  &   \\ [2pt]
       \hline \hline
    \end{tabular*}
\end{table}

\subsection{$^{108}$Xe}

   The N=Z isotope of Xenon has not been experimentally studied yet.  The ISM predictions for the yrast
    band are collected in Table~\ref{tab:xe108y}.  The results at low spin resemble very much to the ones obtained for the
    heavier isotopes, even if the the quadrupole collectivity is slightly smaller.  The backbending  occurs at J=16
   and it is preceded by an upbending at J=14 while in $^{110}$Xe the backbending occurs sharply at J=14.
   We shall dwell with this issue in the next section.  As predicted by the pseudo-SU3 model, $^{108}$Xe
   exhibits also a $\gamma$ band whose properties are listed in Table~\ref{tab:xe108g}. Notice that, in spite
   of some small irregularities, both the yrast and the $\gamma$ band share a common intrinsic state whose
   quadrupole moment is very close to the pseudo-SU3 number. Using equation (1) we can deduce a value 
   $\gamma$=24$^o$, indicating the triaxial nature of this nucleus. This value of $\gamma$ is larger than
   the one obtained for $^{110}$Xe, again in good accord with the model predictions.

\begin{table}
\caption{\label{tab:xe108y} Properties of the yrast band of $^{108}$Xe 
 (energies in MeV, Q's in efm$^2$ and BE2's in e$^2$fm$^4$)}
\begin{tabular*}{\linewidth}{@{\extracolsep{\fill}}cccrrrrr}

\hline \hline \\ [2pt]
J & E* & E$_{\gamma}$ & BE2 & Q$_{sp}$ & Q$_0$      & Q$_0$  & $\beta$\\
  &    &              &       &       & (BE2)    & (Q$_{sp}$) \\ [2pt]  \hline \\ [2pt]
 2$^+$  & 0.41 & 0.41 & 888  & -57  & 200 & 211 &  0.16 \\ [2pt]
 4$^+$  & 1.03 & 0.62 & 1285 & -71  & 195 & 210 &  0.16 \\ [2pt]
 6$^+$  & 1.89 & 0.86 & 1345 & -65  & 163 & 208 &  0.16 \\ [2pt]
 8$^+$  & 2.90 & 1.01 & 1404 & -64  & 154 & 206 &  0.16 \\ [2pt]
10$^+$  & 4.03 & 1.13 & 1334 & -67  & 160 & 198 &  0.15 \\ [2pt]
12$^+$  & 5.37 & 1.34 & 1129 & -71  & 175 & 182 &  0.15 \\ [2pt]
14$^+$  & 6.69 & 1.32 & 990  & -79  & 176 & 168 &   0.14    \\ [2pt]
16$^+$  & 7.75 & 1.07 & 0.1  & -137 &     &     &       \\ [2pt]
18$^+$  & 8.34 & 0.58 & 830  & -140 &     &     &       \\ [2pt]
20$^+$  & 9.24 & 0.90 & 753  & -143 &     &     &       \\ [5pt]
       \hline \hline
    \end{tabular*}
\end{table}

\begin{table}
\caption{Properties of the $\gamma$ band of $^{108}$Xe
 (energies in MeV, Q's in efm$^2$ and BE2's in e$^2$fm$^2$)}
\label{tab:xe108g}
\begin{tabular*}{\linewidth}{@{\extracolsep{\fill}}cccrrrr}
    \hline \hline \\ [2pt]
J & E* & E$_{\gamma}$ & BE2 & Q$_{sp}$      & Q$_0$(Q$_{sp}$)  & $\beta$\\ [2pt]
\hline \\ [2pt]
2$^+_2$ & 1.03 &      &      & +59  &   196      &  0.16      \\ [2pt]
3$^+$   & 1.28 & 0.25 & 1624 & -1.3 &          &       \\ [2pt]
4$^+_2$ & 1.51 & 0.24 & 1090 & -38  &  265 &  0.18 \\ [2pt]
5$^+$   & 1.84 & 0.32 & 882  & -51  & 220 &  0.17 \\ [2pt]
6$^+_2$ & 2.25 & 0.42 & 372  & -83   & 290 &  0.19 \\ [5pt]
       \hline \hline
    \end{tabular*}
\end{table}

\subsection{The effect of the excitations of the  $^{100}$Sn  core}

 At the very origin of this paper was the question about the stiffness of the $^{100}$Sn  core
 and the eventual need of a soft core in order to understand the collective aspects found in the
 very proton rich Xenon isotopes. We have argued "in extenso" that collectivity can be obtained
 without any opening of the doubly magic $^{100}$Sn  core. However, it doesn't exist in nature such
 a thing as a perfectly closed core. The dimensions of the basis make it impossible to perform
 calculations adding the 0g$_{9/2}$ orbit to the r4h valence space. Thus we have proceeded as follows:
 In the first place we have repeated the calculations in the r4 space --{\it i.e.}  removing the 0h$_{11/2}$ orbit--
 The quadrupole properties of the low lying states--the ones we are after- remain unchanged in all cases. 
 For the lower part of the yrast band, the effect of this removal is just an increase of the moments of inertia
 by a factor two. Correspondingly, the spectra are much more compressed. In the next set of calculations we 
 allow 1p-1h and 2p-2h excitations from the  0g$_{9/2}$ orbit into the r4 space. Even in this case the
 dimensions are huge (5 x 10$^9$ for the 2p-2h calculation in  $^{110}$Xe), and  in fact the 2p-2h calculation for
 $^{112}$Xe is out of reach. In these calculations  $^{100}$Sn is still a very good doubly closed shell. At the
 1p-1h level only  0.25  particles (out of 20) are promoted to the r4 orbits. At the 2p-2h level  0.5 particles
 are promoted. In spite of the small size of the vacancy, its effect in the E2 properties is not negligible
  at all.  The 1p-1h excitations
 are the ones that are more efficient in the building up of the quadrupole collectivity, leading to  increases
  of the intrinsic quadrupole moments of 15\%,
 17\% and 20\% for  $^{108}$Xe,  $^{110}$Xe and $^{112}$Xe,  respectively. Adding the 2p-2h excitations does not
 change these numbers a lot, they rise to 19\% and  21\% for  $^{108}$Xe and  $^{110}$Xe.  Notice that these 
 percentages have to be doubled to get the effect on  the transition probabilities.
 
 \subsection{Backbending and alignment}
 
\begin{figure}[h]
 \begin{center}
   \includegraphics[width=0.5\textwidth]{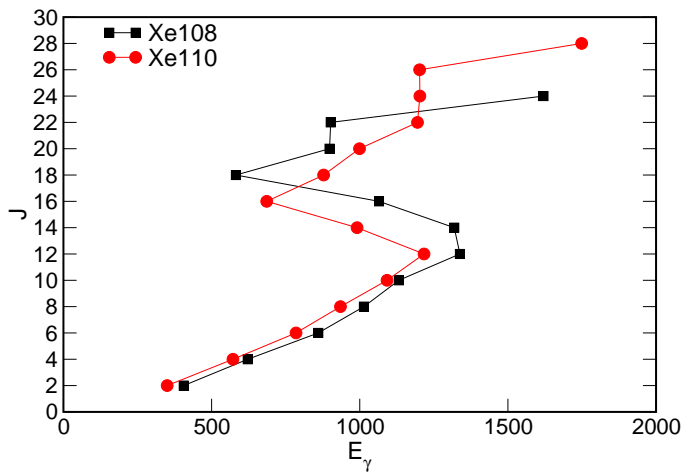}
\caption{\label{fig:xe108-110bb}(color online) Backbending plot of the theoretical yrast bands of  $^{108}$Xe
and $^{110}$Xe}
  \end{center}
\end{figure}

   We have represented the energies of the E2 $\gamma$ cascade along the yrast bands of  $^{108}$Xe
and $^{110}$Xe in the form of a backbending plot in Fig. \ref{fig:xe108-110bb}. We can observe a very similar
(collective) behavior  up to the backbending that occurs at J=14 in both cases. The differences are related to 
the  alignment mechanisms that produce it, whose nature we will explore now. In order to do so we have constructed operators
that count the number of  0h$_{11/2}$ pairs coupled to J=11 T=0 and to J=10 T=1. In the first case, the particles that
align must be one neutron and one proton, while in the second mode they can be two neutrons or two protons as well.    
Let's start with  $^{108}$Xe.  
\begin{figure}[h]
 \begin{center}
    \includegraphics[width=0.5\textwidth]{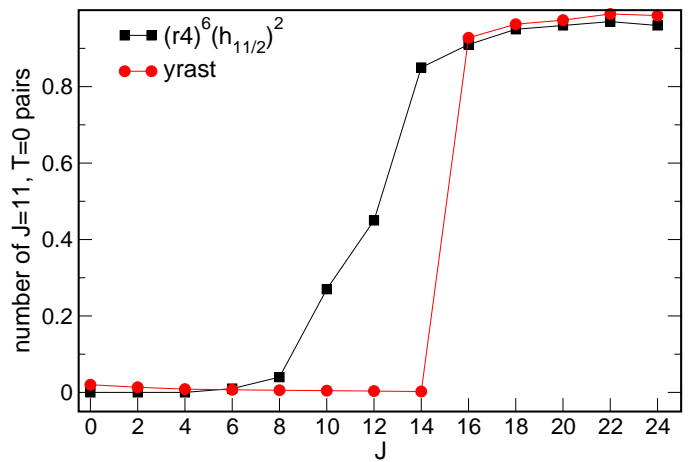}
\caption{\label{fig:xe108alin}(color online) The number of J=11 T=0 pairs in the yrast band of $^{108}$Xe. The same for the
 yrast states of the configuration (r4)$^6$(h$_{11/2}$)$^2$ }
  \end{center}
\end{figure}
 The number of J=11 T=0 neutron proton pairs along the yrast band  is shown in
Fig. \ref{fig:xe108alin}. We observe a sudden transition from zero aligned pairs up to J=14,  to nearly one fully aligned pair
at J=16. From there,  up to the band termination, the number of aligned pairs approaches slowly one.  This reflects in the backbending
in a subtle way. As we can see the real backbending in this case is preceded by an upbending which does not involve
alignment but is probably due to the mixing associated to the band crossing.
 In the same figure we have plotted
the yrast band of the configuration that has two particles blocked in the 0h$_{11/2}$ orbit.
Here the alignment is much smoother. Notice that  only when this configuration is fully
aligned it  becomes the physical yrast band. 
\begin{figure}[h]
 \begin{center}
    \includegraphics[width=0.5\textwidth]{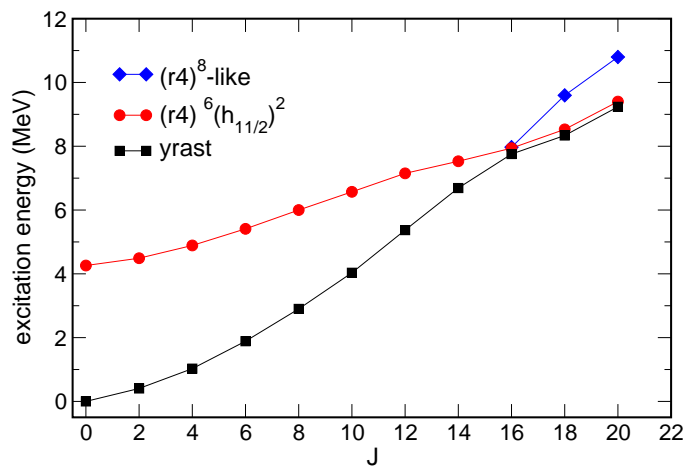}
\caption{\label{fig:xe108cros}(color online) The yrast band of  $^{108}$Xe. Also plotted the
yrast states of the configuration (r4)$^6$(h$_{11/2}$)$^2$ (circles) and the lowest (r4)$^8$-like
states beyond J=16 (lozenges) to illustrate the origin of the backbending}
  \end{center}
\end{figure}
This is better seen in Fig.~\ref{fig:xe108cros} which is alike to
a cartoon of the band crossing.  At low spin the yrast band is dominated by the  (r4)$^8$ configurations, with
the (r4)$^6$(h$_{11/2}$)$^2$ ones lying at about 4~MeV. The crossing happens at J=16 producing local distortions
in the backbending plot. We have been able to locate the states belonging to the (r4)$^8$  yrast band beyond the crossing point. Once the crossing has taken place, the backbending disappears.
\begin{figure}[h]
 \begin{center}
    \includegraphics[width=0.5\textwidth]{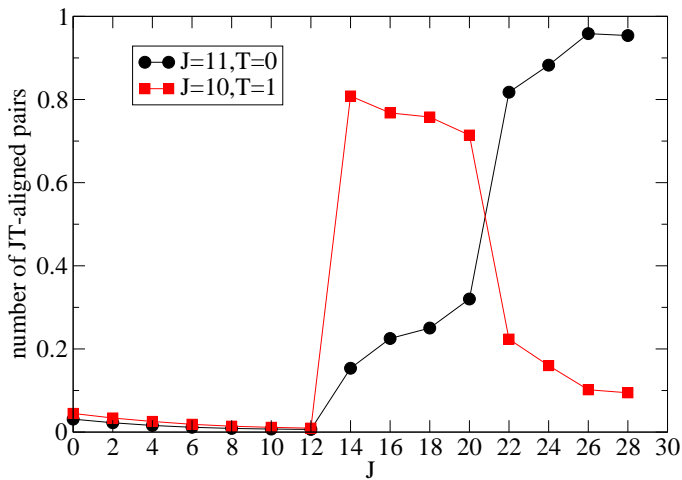}
\caption{\label{fig:xe110alin}(color online) The number of J=10 T=1 and J=11 T=0 pairs in the yrast band of 
$^{110}$Xe}
  \end{center}
\end{figure}
The situation is quite different in the case of the N$\ne$Z nucleus $^{110}$Xe as can be gathered from
 Fig.~\ref{fig:xe110alin}. Here the backbending occurs abruptly at J=14 and corresponds to the alignment
 of (mostly) two neutrons to J=10 T=1. Beyond J=16 the yrast line becomes parallel to the low energy one, as if 
 the angular momentum put in the system  were again of collective nature. At J=22 the isovector alignment
 is depressed and the isoscalar one takes over till the band termination producing new irregularities in the
 backbending plot.

\section{The role of the isoscalar and isovector pairing}

\begin{figure}[h]
 \begin{center}
   \includegraphics[width=0.5\textwidth]{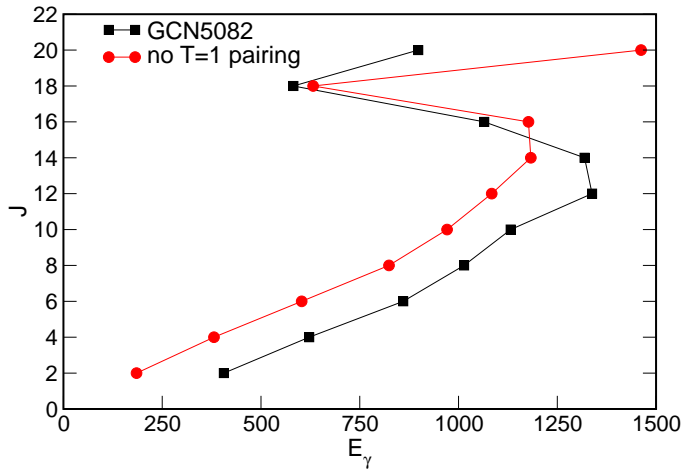}
\caption{\label{fig:xe108-T1}(color online) Backbending plot of the yrast band of  $^{108}$Xe. Results with the
GCN5082 interaction before (squares) and after (circles) removing the isovector J=0 pairing}
  \end{center}
\end{figure}

Quite some time ago, the question of the pairing modes near N=Z  knew an uprise
of interest as more and more experimental data accumulated on this class of nuclei.
Among the topics of interest were  the Wigner energy and the possibility of
finding manifestations of an isoscalar pairing condensate (deuteron like) in the
yrast bands of these proton rich nuclei in the form of delayed alignments etc. 
  \cite{sade97,dopi98}.  It was realized relatively soon that these effect were
 bound to be elusive, because, in normal circumstances, the spin orbit splitting hinders
 strongly the deuteron-like condensation \cite{Poves1998}.
Recent studies of the beta-decay of $^{62}$Ge, where
super-allowed beta-transitions due to the 
existence of the T=0 collective 1$^+$ state 
were predicted \cite{ge62gt}, have ruled out such a possibility as well.

On the contrary, the effect of the isovector pairing channel in the rotational properties of
the N$\sim$Z is well understood. It is directly linked to the moment of inertia of the band,
and does not affect appreciably its electromagnetic decay properties. We have verified that
for the light Xenon isotopes. In Fig.~\ref{fig:xe108-T1} we compare the energies of the $\gamma$ 
cascades along the yrast band for $^{108}$Xe using the effective interaction GCN5082 on one
side and, on the other, this same interaction after removal of an schematic isovector pairing hamiltonian
whose coupling strength is obtained as in ref. \cite{duzu96}. Below the backbending the effect is just
an increase of the moment of inertia by a factor two. As expected, the E2 properties do not change.
The backbending is delayed two units of angular momentum.
\vspace*{0.5cm}

\begin{figure}[h]
 \begin{center}
    \includegraphics[width=0.5\textwidth]{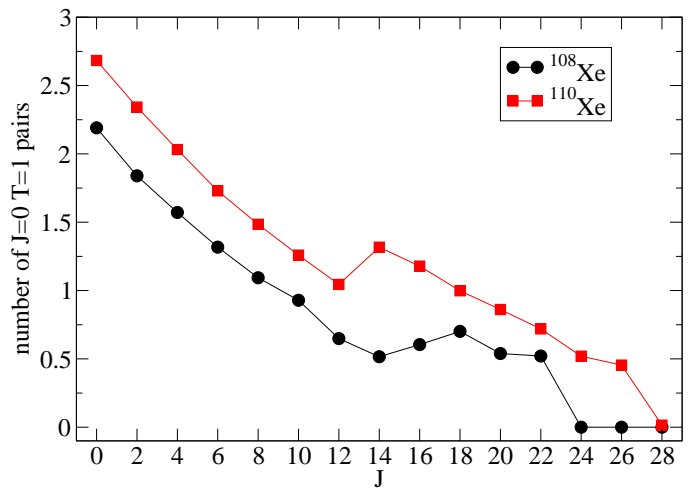}
\caption{\label{fig:xe108-110-T1}(color online) The number of J=0 T=1 pairs in the yrast bands of $^{108}$Xe
and $^{110}$Xe.}
  \end{center}
\end{figure}

\begin{figure}[h]
 \begin{center}
    \includegraphics[width=0.5\textwidth]{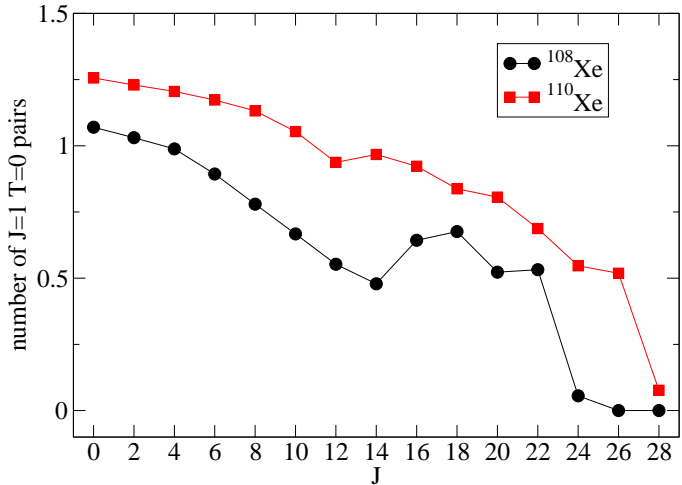}
\caption{\label{fig:xe108-110-T0}(color online) The number of J=1 T=0 pairs in the yrast bands of $^{108}$Xe
and $^{110}$Xe.}
  \end{center}
\end{figure}

To verify the possibility of existence of pair condensates
in the ground states of Xenon 
isotopes we have also constructed pair counting operators, 
and have computed their expectation values along their  yrast 
bands. The results for the isovector pairs are plotted in Fig.~\ref{fig:xe108-110-T1} .
A fully condensed state should have nearly four pairs for  $^{108}$Xe and
five pairs in $^{110}$Xe. What we find for the ground states, which is consistent in both cases, 
amounts to one half
of the value expected for the condensate, indeed a quite important pairing contribution.
As the angular momentum increases, the pair content decreases linearly reaching negligible values
at the backbending. The pattern in this region is simpler in  $^{108}$Xe  than  $^{108}$Xe, due to
the different alignment mechanisms in both isotopes which we have discussed already. 
The results for the isocalar pairs are plotted in Fig.~\ref{fig:xe108-110-T0}. In the limit of an isovector
condensate we should expect typically four pairs, and we can see that we are far from that. In addition,
the "a priori" more favorable case, N=Z, is depressed with respect to the N$\ne$Z. All in all
there seems to be no indications of any structural effect due to the isoscalar pairing channel. 

\section{Conclusions} We have carried out large scale shell model calculations for the lighter
Xenon isotopes in the valence space
r4h that encompasses all the orbits between the magic numbers 50 and 82.  We obtain collective
behaviors of triaxial nature and can explain the experimental results withouts resorting to large openings of the
$^{100}$Sn core. We propose a mechanism that can explain the very large values of the intrinsic quadrupole moments of the Xenon isotopes at mid neutron shell based in variants of the SU3 symmetry. We have shown that the 
backbending  in   $^{108}$Xe is produced by the alignment of a neutron proton pair  in the 0h$_{11/2}$
orbit to the maximum allowed spin J=11. In $^{110}$Xe it is a two step process, first a pair of neutrons 
align to J=10 and afterwords the neutron proton alignment takes over. Finally we have studied the pair
content of the yrast states. Isovector J=0 pairs have a large presence in the lowest states of the yrast
bands of  $^{108}$Xe and  $^{110}$Xe. On the contrary, the deuteron like J=1 isoscalar pairs have a 
negligible presence in these nuclei.

{\bf Acknowledgments.} This work is partly supported  by a grant of
                       the Spanish Ministry of Education and Science
                       (FPA2009-13377), by the IN2P3(France)-CICyT(Spain)
                       collaboration agreements, by the Spanish
                       Consolider-Ingenio 2010 Program CPAN
                       (CSD2007-00042) and by the Comunidad de Madrid
                       (Spain), project HEPHACOS S2009/ESP-1473.

\end{document}